\documentclass[aps,prl,twocolumn,superscriptaddress]{revtex4-1}

\usepackage{amsfonts}
\usepackage{amssymb}
\usepackage{amsmath}
\usepackage{upgreek}
\usepackage{color}
\usepackage{graphicx}
\usepackage[colorlinks=true,citecolor=blue,linkcolor=blue,urlcolor=blue]{hyperref}

\newcommand{\beginsupplement}{%
        \setcounter{table}{0}
        \renewcommand{\thetable}{S\arabic{table}}%
        \setcounter{figure}{0}
        \renewcommand{\thefigure}{S\arabic{figure}}%
        \setcounter{equation}{0}
        \renewcommand{\theequation}{S\arabic{equation}}%
     }
\newcounter{SN}
\newcommand{\rSN}[1]{\refstepcounter{SN}\label{#1}}

\newcommand{\MIT}{Massachusetts Institute of Technology, Department of Physics, Cambridge, Massachusetts 02139, USA.}
\newcommand{\Geballe}{Geballe Laboratory for Advanced Materials, Stanford University, Stanford, California 94305, USA.}
\newcommand{\StanfordAP}{Department of Applied Physics, Stanford University, Stanford, California 94305, USA.}
\newcommand{\StanfordMSE}{Department of Materials Science and Engineering, Stanford University, Stanford, California 94305, USA.}
\newcommand{\SIMES}{SIMES, SLAC National Accelerator Laboratory, Menlo Park, California 94025, USA.}
\newcommand{\APS}{Advanced Photon Source, Argonne National Laboratory, Argonne, Illinois, 60439, USA.}
\newcommand{\SLAC}{SLAC National Accelerator Laboratory, Menlo Park, California 94025, USA.}

\newcommand{\DESY}{Center for Free-Electron Laser Science, DESY, Notkestra{\ss}e 85, 22607 Hamburg, Germany.}
\newcommand{\SYSU}{School of Electronics and Information Technology, Sun Yat-sen University, Guangzhou, Guangdong 510006, China.}
\newcommand{\Harvard}{Department of Physics, Harvard University, Cambridge, Massachusetts, 02138, USA.}

\begin{document}

\title{Light-Induced Charge Density Wave in LaTe$_3$}

\author{Anshul~Kogar}
\thanks{These authors contributed equally to this work.}
\affiliation{\MIT}
\author{Alfred~Zong}
\thanks{These authors contributed equally to this work.}
\affiliation{\MIT}
\author{Pavel~E.~Dolgirev}
\affiliation{\Harvard}
\author{Xiaozhe~Shen}
\affiliation{\SLAC}
\author{Joshua~Straquadine}
\affiliation{\Geballe}
\affiliation{\StanfordAP}
\affiliation{\SIMES}
\author{Ya-Qing~Bie}
\thanks{Present address: \SYSU}
\affiliation{\MIT}
\author{Xirui~Wang}
\affiliation{\MIT}
\author{Timm~Rohwer}
\thanks{Present address: \DESY}
\affiliation{\MIT}
\author{I-Cheng~Tung}
\affiliation{\APS}
\author{Yafang~Yang}
\affiliation{\MIT}
\author{Renkai~Li}
\affiliation{\SLAC}
\author{Jie~Yang}
\affiliation{\SLAC}
\author{Stephen Weathersby}
\affiliation{\SLAC}
\author{Suji~Park}
\affiliation{\SLAC}
\affiliation{\StanfordMSE}
\author{Michael~E.~Kozina}
\affiliation{\SLAC}
\author{Edbert~J.~Sie}
\affiliation{\Geballe}
\affiliation{\SIMES}
\author{Haidan~Wen}
\affiliation{\APS}
\author{Pablo~Jarillo-Herrero}
\affiliation{\MIT}
\author{Ian~R.~Fisher}
\affiliation{\Geballe}
\affiliation{\StanfordAP}
\affiliation{\SIMES}
\author{Xijie~Wang}
\affiliation{\SLAC}
\author{Nuh~Gedik}
\email[Correspondence to: ]{gedik@mit.edu}
\affiliation{\MIT}
\begin{abstract}
When electrons in a solid are excited with light, they can alter the free energy landscape and access phases of matter that are beyond reach in thermal equilibrium. This accessibility becomes of vast importance in the presence of phase competition, when one state of matter is preferred over another by only a small energy scale that, in principle, is surmountable by light. Here, we study a layered compound, LaTe$_3$, where a small in-plane (\textit{a}-\textit{c} plane) lattice anisotropy results in a unidirectional charge density wave (CDW) along the \textit{c}-axis. Using ultrafast electron diffraction, we find that after photoexcitation, the CDW along the \textit{c}-axis is weakened and subsequently, a different competing CDW along the \textit{a}-axis emerges. The timescales characterizing the relaxation of this new CDW and the reestablishment of the original CDW are nearly identical, which points towards a strong competition between the two orders. The new density wave represents a transient non-equilibrium phase of matter with no equilibrium counterpart, and this study thus provides a framework for unleashing similar states of matter that are ``trapped" under equilibrium conditions.
\end{abstract}
\date{March 23, 2019}
\maketitle

A major theme in condensed matter physics is the relationship between proximal phases of matter, where one ordered ground state gives way to another as a function of some external parameter such as pressure, magnetic field, doping, or disorder. It is in such a neighborhood that we find colossal magnetoresistance in manganites \cite{Tokura2006} and unconventional superconductivity in heavy fermion, copper oxide, and iron-based compounds \cite{Norman2011}. In these materials, the nearby ground states can affect one another in several ways. For example, phases can compete, impeding the formation of one state in place of another. This scenario is played out, for instance, in La$_{2-x}$Ba$_x$CuO$_4$ at $x=1/8$, where the development of alternating charge and spin-ordered regions prevents the onset of superconductivity \cite{Abbamonte2005, Tranquada2013}. On the other hand, fluctuations of an adjacent phase can help another be realized, such as in $^3$He, where ferromagnetic spin fluctuations enable the atoms to form Cooper pairs and hence a $p$-wave superfluid \cite{Norman2011, Leggett1975}. In more complicated situations, such as in manganites, nanoscale phase separation occurs, where local insulating antiferromagnetism coexists next to patches of metallic ferromagnetism, resulting in large magnetic and electrical responses to small perturbations \cite{Tokura2006}. In each case, the macroscopic properties of a material are heavily influenced by the nearby presence of different phases.

Intense light pulses have recently emerged as a tool to tune between neighboring broken-symmetry phases of matter \cite{Nova2018, Li2018, Fausti2011, Tokura2007, Nasu2004, Sie2019}. Conventionally, light pulses are used to restore symmetry, but in certain cases symmetries can also be broken. For instance, exposing SrTiO$_3$ to midinfrared radiation has led to ferroelectricity \cite{Nova2018, Li2018}, while ferromagnetism has been induced in a manganite with near-infrared light \cite{Tokura2007}. In this Letter, we examine a quasi-two-dimensional material, LaTe$_3$, where a unidirectional CDW phase is only present along the $c$-axis with no counterpart along the nearly-equivalent, perpendicular $a$-axis. We show that femtosecond light pulses can be used to break translational symmetry and unleash an $a$-axis CDW. Using ultrafast electron diffraction (UED), we visualize this process and track both order parameters simultaneously, gaining a unique perspective of both orders in the time domain.

\begin{figure*}[htb!]
	\includegraphics[scale=1.11]{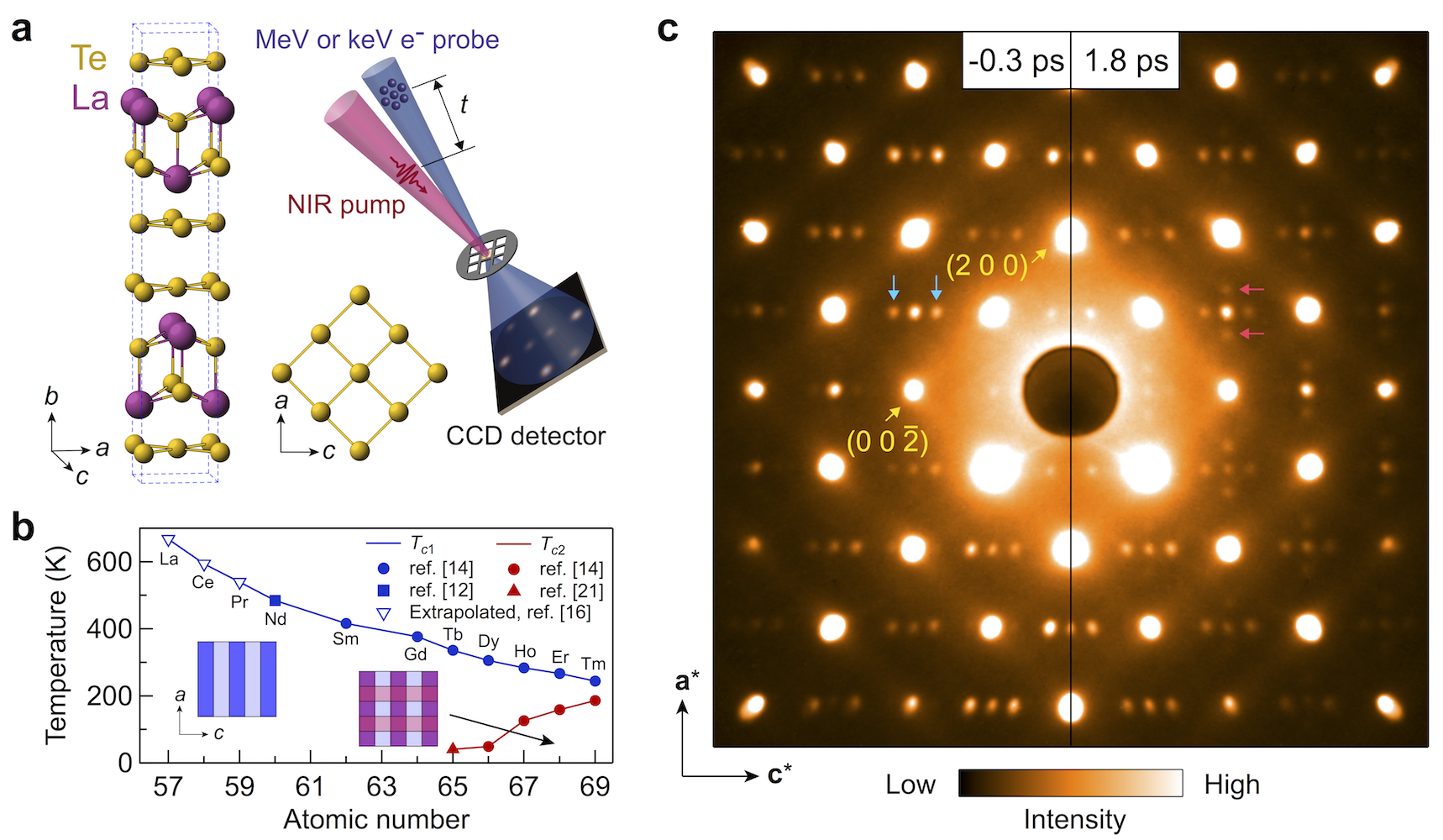}
	\caption{\textbf{Observation of a transient CDW induced by an 80-fs, 800-nm laser pulse.} \textbf{a},~\emph{Left}: schematic of the LaTe$_3$ crystal structure, where dashed lines indicate the unit cell. Electronic states near the Fermi surface arise from the planar Te sheets, which are nearly square-shaped with a slight in-plane anisotropy ($a=0.997c$) at room temperature \cite{Malliakas2006,Ru2008thesis}. \emph{Right}:  schematic of the ultrafast electron diffraction setup in the transmission mode. Both 26~keV and 3.1~MeV electrons were used (see Methods and Supplementary Note~\ref{sn:keV}). Exfoliated samples were mounted on a 10\,nm-thick silicon nitride window. \textbf{b},~Summary of the two CDW transition temperatures, $T_{c1}$ and $T_{c2}$, across the rare-earth series. Insets are schematics of the unidirectional CDW below $T_{c1}$ and bidirectional CDW below $T_{c2}$, respectively. \textbf{c},~Electron diffraction patterns before (left) and 1.8\,ps after (right) photoexcitation with a near-infrared (NIR) laser pulse, taken at 3.1\,MeV electron kinetic energy. Blue and red arrows indicate the equilibrium CDW peaks along the $c$-axis and the light-induced CDW peaks along the $a$-axis, respectively. The right half is a mirror reflection of the left half, so they denote the same set of peaks in the diffraction pattern. A full diffraction image is shown in Fig.\,\ref{fig:keV}(a). $\mathbf{a^*}\equiv(2\pi/|\mathbf{a}|)\hat{\mathbf{a}}$ and $\mathbf{c^*}\equiv(2\pi/|\mathbf{c}|)\hat{\mathbf{c}}$ are reciprocal lattice unit vectors.}
\label{fig:intro}
\end{figure*}

LaTe$_3$ is a member of the rare-earth tritellurides ($R$Te$_3$, where $R$ denotes a rare-earth element). These materials possess a layered, quasi-tetragonal structure (Fig.\,\ref{fig:intro}(a)) with a slight in-plane anisotropy ($a\geq0.997c$, Fig.\,\ref{fig:ac_lattice}(b)) \cite{Malliakas2006,Ru2008thesis}, which leads to a preferred direction for the CDW order along the $c$-axis. Depending on the specific rare-earth element, some of the members display a CDW only along the $c$-direction while others have an additional CDW along the orthogonal $a$-direction (Fig.\,\ref{fig:intro}(b)). All of them  share a similar normal-state Fermi surface that arises from the nearly square-shaped Te sheets, and the rare-earth atoms, with different radii, effectively serve to apply chemical pressure \cite{Ru2008, DiMasi1995}. As one moves from lighter to heavier rare-earth elements, the transition temperature of the CDW along the $c$-axis, $T_{c1}$, decreases while that along the $a$-axis, $T_{c2}$, is first finite in TbTe$_3$ and increases with atomic number (Fig.\,\ref{fig:intro}(b)). This relationship strongly suggests that the two CDWs compete in equilibrium. In the material we study here, LaTe$_3$, $T_{c1}$ is estimated to be $\sim670$\,K \cite{Hu2014}, and a CDW along the $a$-axis does not exist.

\begin{figure*}[htb!]
	\includegraphics[scale=0.56]{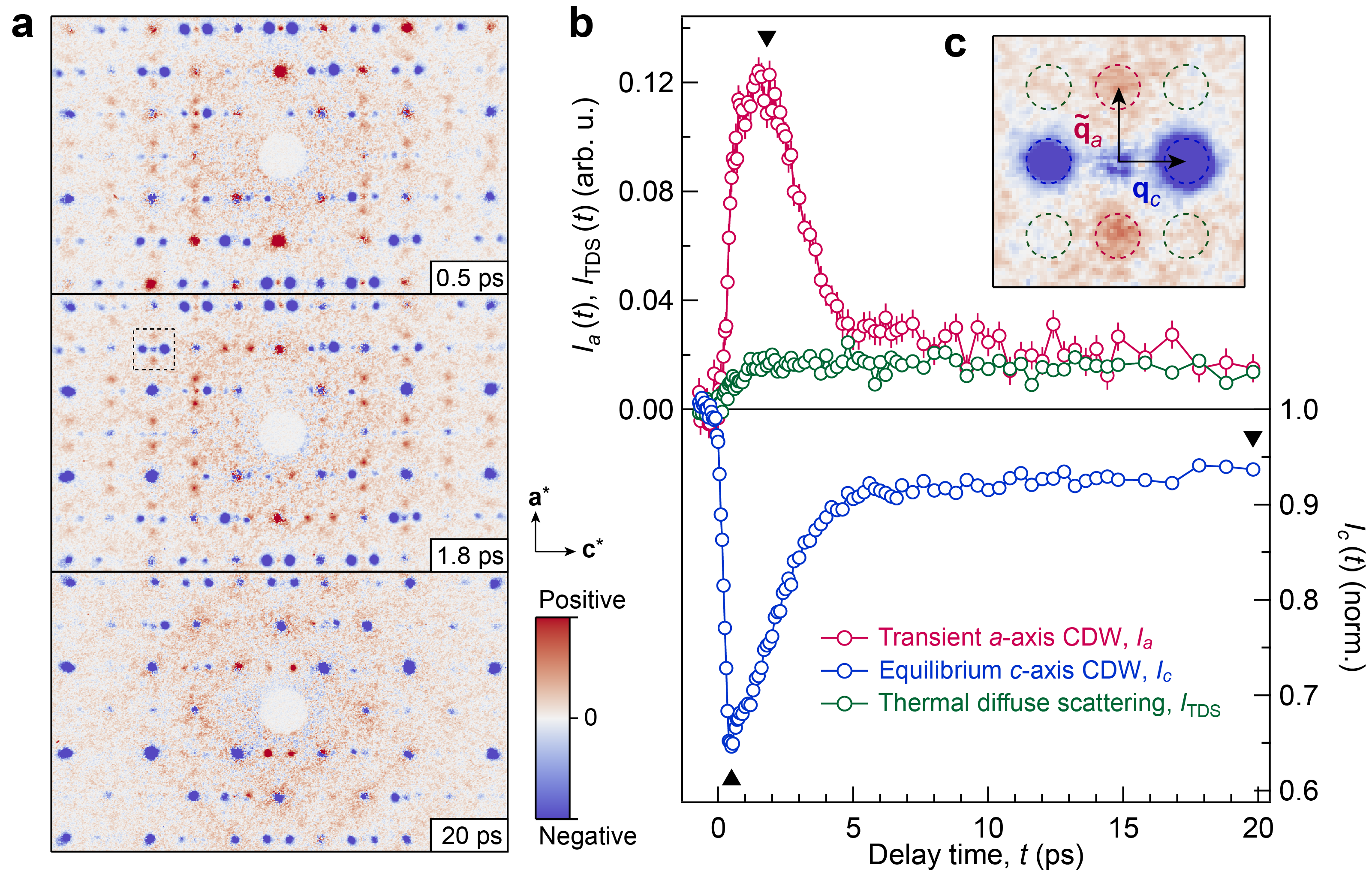}
	\caption{\textbf{Dynamics of the light-induced CDW.} \textbf{a},~Change in intensities with respect to the diffraction pattern before photoexcitation. Snapshots are taken at three selected pump-probe time delays, as indicated by the triangles in \textbf{b}. \textbf{b},~Time evolution of integrated intensities of the equilibrium $c$-axis CDW peak ($I_c$), the transient $a$-axis CDW peak ($I_a$), and the thermal diffuse scattering ($I_\text{TDS}$). Integration areas are marked by circles in \textbf{c} with corresponding colors. Peaks in multiple Brillouin zones are averaged for improved signal-to-noise ratio. $I_a$ and $I_\text{TDS}$ are vertically offset to have their values zeroed prior to photoexcitation. $I_c$ is normalized by its value before photoexcitation. Error bars represent the standard deviation of noise for $t<0$. \textbf{c},~A zoomed-in view of the dashed square in \textbf{a} at $t=1.8$\,ps. Each integration region has a diameter equal to 1.5 times the full-width-at-half-maximum (FWHM) of the equilibrium CDW peak. The incident pump laser fluence for all panels was 1.3\,mJ/cm$^2$.}
\label{fig:timetrace}
\end{figure*}

To follow the temporal evolution of the CDW after light excitation, we used transmission ultrafast electron diffraction (Fig.\,\ref{fig:intro}(a)), which allows us to capture the ($H$ 0 $L$) plane, with ($H$ $K$ $L$) denoting the Miller indices. In the left panel of Fig.\,\ref{fig:intro}(c), we show a static diffraction pattern of LaTe$_3$ taken before the arrival of the pump laser pulse, where satellite peaks (blue arrows) flanking the main Bragg peaks are observed only along the $c$-axis. These peaks are due to the existence of the equilibrium CDW. In the right panel, we show the diffraction pattern 1.8\,ps after photoexcitation by an 80-fs, 800-nm (1.55-eV) laser pulse, which creates excitations across the single-particle gap and suppresses the CDW along the $c$-axis. As the equilibrium CDW is weakened, new peaks emerge along the $a$-direction (red arrows) independent of the pump laser polarization, a change that can also be visualized in the differential intensity plot in Fig.\,\ref{fig:timetrace}(a). Here, the appearance of a new lattice periodicity along the $a$-axis is clear and we interpret these peaks as signalling the emergence of an out-of-equilibrium CDW. This observation was replicated in four different samples at two separate UED setups, which use 3.1\,MeV and 26\,keV electron kinetic energies respectively (see Methods and Supplementary Note~\ref{sn:keV}).

This non-equilibrium CDW is ephemeral and only lasts for a few picoseconds. In Fig.\,\ref{fig:timetrace}(b), we show the temporal evolution of the integrated intensity of the peaks along both the $a$- and $c$-axis. The intensity of the $a$-axis CDW peak reaches a maximum around 1.8\,ps and then relaxes over the next couple of picoseconds to a quasi-equilibrium value. The residual intensity at long time delays is due to laser pulse-induced heating that causes a thermal occupation of phonons, which is shown in the diffuse scattering trace in Fig.\,\ref{fig:timetrace}(b) and as the overall red background in Fig.\,\ref{fig:timetrace}(a,c). The intensity of the $c$-axis CDW peak shows the opposite behavior: it first reaches a minimum around 0.5\,ps before recovering to a quasi-equilibrium. The initial decay of the $c$-axis CDW occurs markedly faster than the rise of the transient CDW. This is because the suppression of the equilibrium CDW involves a coherent motion of the lattice ions, whose timescale is tied to the period of the CDW amplitude mode \cite{Hellmann2012, Schmitt2008}. On the other hand, incoherent fluctuations dictate the ordering of the $a$-axis CDW, which occurs on a slower timescale (see Supplementary Note~\ref{sn:rise_time}).

\begin{figure*}[htb!]
	\includegraphics[scale=0.56]{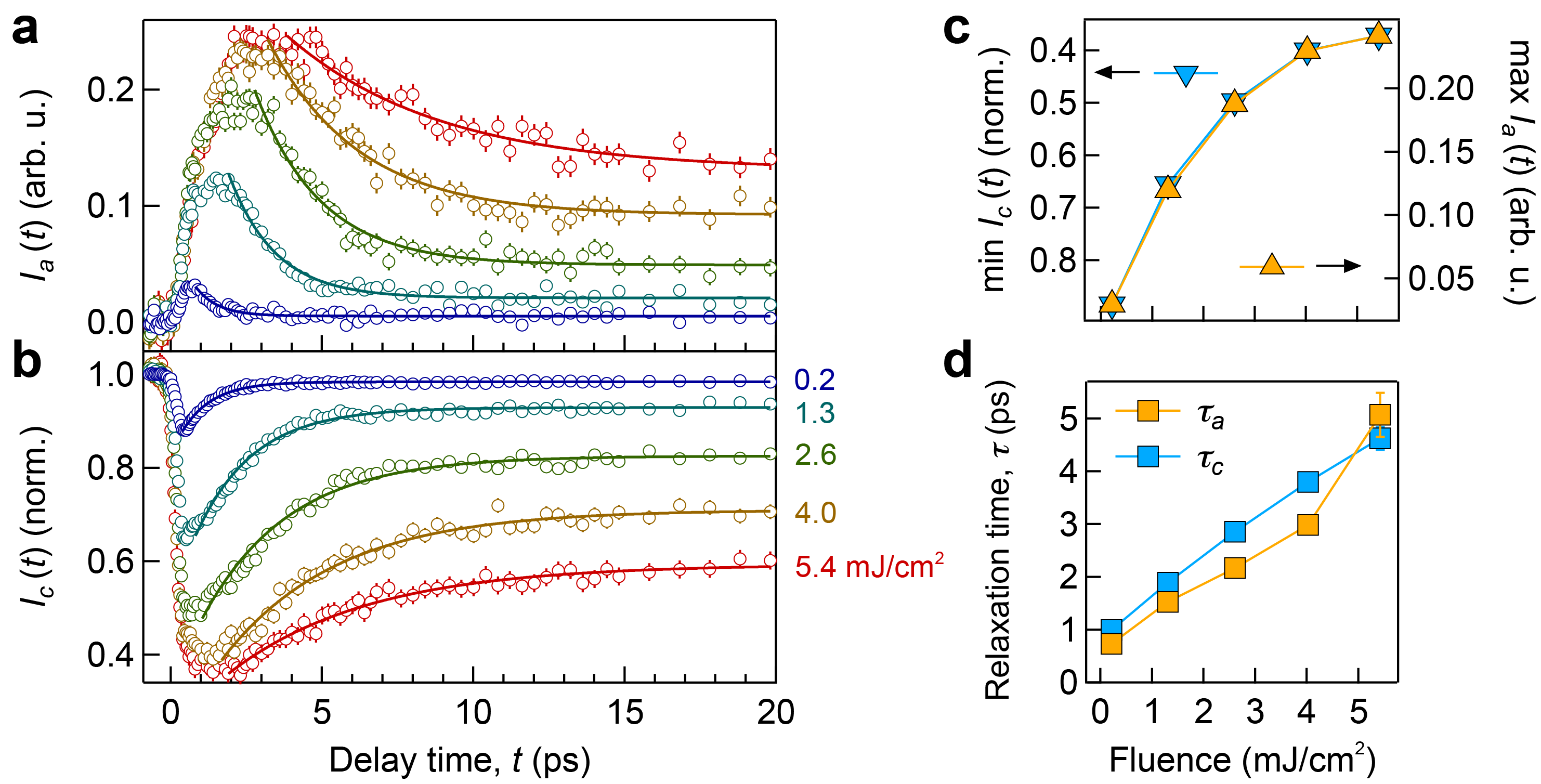}
	\caption{\textbf{Dependence of equilibrium and transient CDW peaks on pump laser fluence}. \textbf{a-b},~Time evolution of integrated intensities for the transient $a$-axis CDW peaks and the equilibrium $c$-axis CDW peaks, respectively. Each color denotes an incident fluence. Error bars are obtained from the standard deviation of noise prior to photoexcitation. Curves are single-exponential fits to the relaxation dynamics. In \textbf{b}, the intensity $I_c$ does not transiently reach zero at high fluence because of background intensities in the diffraction pattern and non-uniform illumination of all layers  of  the  sample  due  to  a  shorter  pump  laser  penetration depth (44\,nm at 800\,nm wavelength) compared to the sample thickness ($\lesssim60$\,nm) used in this case. \textbf{c},~\emph{Left}: minimum value of the integrated intensity for the equilibrium $c$-axis CDW peaks. \emph{Right}: maximum value of the integrated intensity for the transient $a$-axis CDW peaks. \textbf{d},~Characteristic relaxation times at different fluences for the recovery of the $c$-axis CDW peaks ($\tau_c$) and the disappearance of the $a$-axis CDW peaks ($\tau_a$). Error bars, if larger than the symbol size, denote one standard deviation in the corresponding single-exponential fits in \textbf{a} and \textbf{b}.}
\label{fig:fludepend}
\end{figure*}

Despite the disparity in the initial timescales, the relaxation times are nearly identical and the overall intensity changes are perfectly anti-correlated, which suggest that these latter properties are governed by a single underlying mechanism. Figure~\ref{fig:timetrace}(b) shows that one CDW forms at the cost of the other and the two recover back to quasi-equilibrium simultaneously, which, for this fluence, takes a characteristic time of $\tau_a\approx \tau_c \approx1.7$\,ps. The agreement in both the trends in intensities and the characteristic relaxation times is even more striking when examining the data at different photoexcitation fluences. As shown in Fig.\,\ref{fig:fludepend}(a,b) and summarized in Fig.\,\ref{fig:fludepend}(c,d), for each fluence, the two CDWs reach anti-correlated extremum values and relax in almost perfect correspondence. Such a strong correlation in both the intensities and the relaxation timescales naturally points towards a phase competition in this non-equilibrium context where the transient CDW cannot exist once the equilibrium CDW recovers.

\begin{figure*}[htb!]
	\includegraphics[scale=0.72]{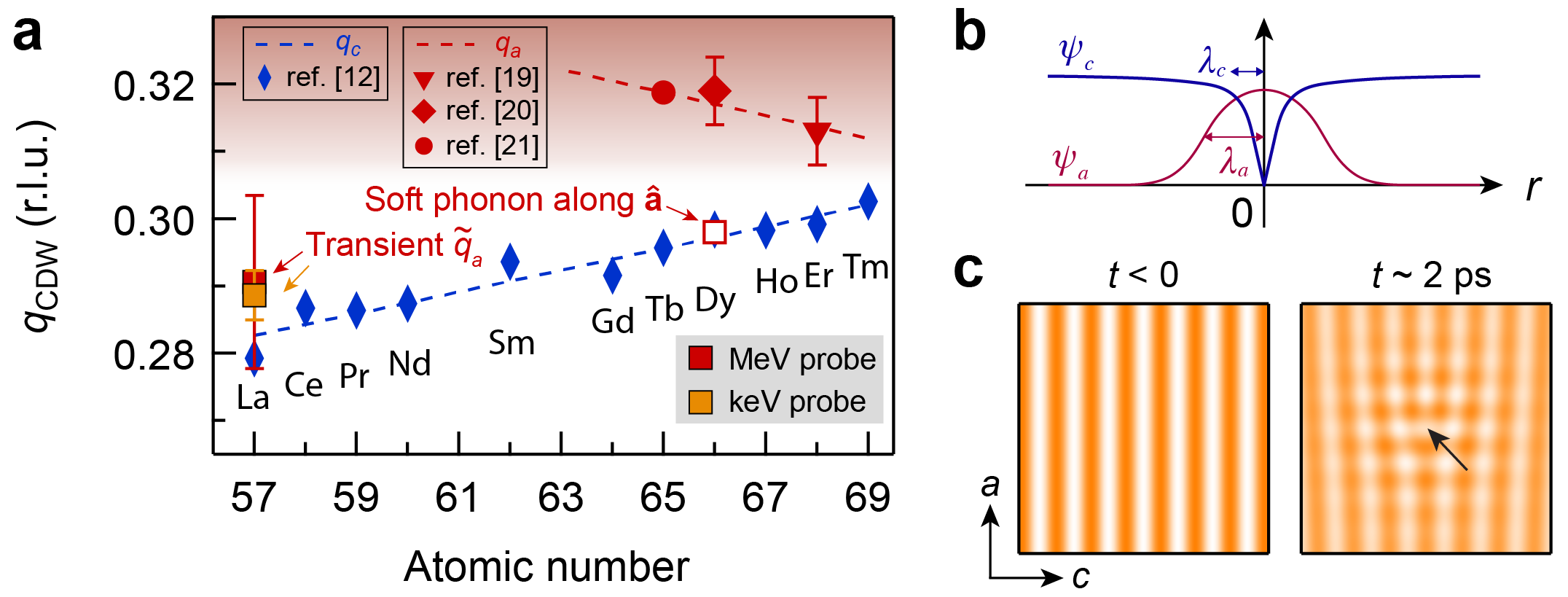}
	\caption{\textbf{Transient CDW seeded by topological defects.} \textbf{a},~Summary of CDW wavevectors across the rare-earth series; r.l.u.: reciprocal lattice unit. Values of $q_c$ (blue diamonds) were taken from Ref.~\cite{Malliakas2006} at the highest temperature below $T_{c1}$. For $q_a$, only TbTe$_3$ \cite{Banerjee2013}, DyTe$_3$ \cite{Maschek2018}, and ErTe$_3$ \cite{Ru2008err} were accurately measured by X-ray diffraction  (see Supplementary Note~\ref{sn:q_cdw} for a discussion on the temperature dependence of $q_c$ and $q_a$ at equilibrium). White square denotes the calculated wavevector of the soft phonon along the $a$-axis, which was confirmed by inelastic X-ray scattering at $T_{c1}$ \cite{Maschek2018}. Red (orange) square denotes the wavevector of the transient $a$-axis CDW, probed by time-resolved MeV (keV) electron diffraction. Dashed lines are guides to eye, highlighting a monotonic trend for both $q_a$ and $q_c$ across the rare-earth elements. Shaded region represents extrapolated values of $q_a$ for light rare-earth elements (La to Gd) if a bidirectional CDW were to form. Error bars, if larger than symbol size, denote reported uncertainty in the literature, or for $\widetilde{q}_a$, the standard deviation of values obtained in multiple Brillouin zones and diffraction images (see Supplementary Note~\ref{sn:q_cdw} for details of determining $\widetilde{q}_a$). \textbf{b},~Schematic of CDW order parameter amplitudes, $\psi_c$ and $\psi_a$, near a topological defect in the equilibrium $c$-axis CDW (see Supplementary Note~\ref{sn:GLtheory}). Characteristic length scales of suppression in $\psi_c$ and enhancement in $\psi_a$ are labeled by $\lambda_c$ and $\lambda_a$, respectively. \textbf{c},~Schematic of charge density waves in real space before (left) and approximately 2\,ps after (right) photoexcitation. Stripe brightness indicates the strength of the CDW amplitude. A dislocation (black arrow) is used as an example of a topological defect in the $c$-axis CDW after photoexcitation.}
\label{fig:cartoon}
\end{figure*}

Upon close scrutiny of the transient CDW wavevector, $\widetilde{q}_a$, it appears that this CDW is a genuinely non-equilibrium phase (we use a tilde to denote the non-equilibrium value). Notably, the wavevector does not resemble values seen in other rare-earth tritellurides that exhibit an equilibrium $a$-axis CDW \cite{Ru2008err, Maschek2018, Banerjee2013}. The transient CDW has an incommensurate wavevector, $\widetilde{q}_a=0.291(13)$, expressed in reciprocal lattice units (see red square in Fig.\,\ref{fig:cartoon}(a) and Supplementary Note~\ref{sn:q_cdw}). On the other hand, the $q_a$ measured in other rare-earth tritellurides in equilibrium are significantly larger (Fig.\,\ref{fig:cartoon}(a)). According to the trend of $q_a$ with rare-earth mass, one would predict an even larger wavevector for LaTe$_3$. Instead, $\widetilde{q}_a$ is closer in value to the markedly smaller wavevector of the $c$-axis CDW. Thus, the observed $\widetilde{q}_a$ of the transient CDW highlights that it is not a trivial extension to an equilibrium $a$-axis CDW.

We can gain some insight into the origin of the anomalous wavevector from previous inelastic X-ray scattering measurements and density functional theory calculations on DyTe$_3$, which is in the same CDW family \cite{Maschek2018}. In DyTe$_3$, when the $c$-axis CDW develops at 308\,K, strong CDW fluctuations are also seen along the $a$-direction in the form of phonon softening, namely, a marked decrease in the phonon frequency. As shown in Fig.\,\ref{fig:cartoon}(a), these fluctuations occur at a wavevector $q_{a\text{,\,soft}}$, which is comparable in magnitude to the $c$-axis CDW, $q_c$. However, when the $a$-axis CDW eventually forms at 50\,K, it does so at a larger wavevector, $q_a$ (i.e., $q_{a\text{,\,soft}}\approx q_c \ll q_a$). Given the negligible $a/c$-anisotropy in the normal-state Fermi surface, the reason for this difference in wavevectors is the following: when the $a$-axis CDW forms at low temperature, it does so after the $c$-axis CDW has already opened a gap at portions of the Fermi surface, which changes the nesting conditions \cite{Moore2010}. Returning to LaTe$_3$, we observe $\widetilde{q}_a \approx q_c$ (see Supplementary Note~\ref{sn:q_cdw}), which suggests that the transient $a$-axis CDW looks more akin to one that would have formed at high temperature had the $c$-axis CDW not prevented it from doing so.

To explain all of these observations within a consistent framework, we propose a picture where the non-equilibrium CDW arises due to the existence of topological defects in the $c$-axis CDW (Fig.\,\ref{fig:cartoon}(c)). The presence of these defects was recently evidenced in LaTe$_3$ upon photoexcitation \cite{Zong2019} and visualized by scanning tunneling microscopy of palladium-intercalated ErTe$_3$ \cite{Fang2019}. In spatial regions where the dominant $c$-axis order is suppressed, such as in topological defects, the sub-dominant $a$-axis phase can develop (Fig.\,\ref{fig:cartoon}(b)) \cite{Arovas1997,Lake2001, Hoffman2002}. The benefit of this picture is that it can explain several observations that are difficult to capture in other theoretical scenarios (see Supplementary Note~\ref{sn:other_theories}). First, the transient CDW forms despite only a partial suppression of the $c$-axis CDW, as shown in Fig.\,\ref{fig:fludepend}(a,b). From equilibrium, we know that any finite $c$-axis CDW amplitude necessarily forbids the presence of an $a$-axis CDW in LaTe$_3$. The presence of topological defects, however, explains this apparent puzzle considering that it allows for the local suppression of the $c$-axis CDW. This local constraint also accounts for the anomalous wavevector, $\widetilde{q}_a$, since the transient CDW nucleates in the absence of the $c$-axis CDW. Furthermore, the coincidence of relaxation timescales is naturally explained in this scenario: as the defects annihilate, the transient CDW can no longer be sustained and the equilibrium $c$-axis CDW necessarily recovers \cite{Zong2019}.

To place the proposed mechanism on a firmer theoretical footing, we have performed a Ginzburg-Landau analysis in two-dimensional space involving two complex order parameters, $\psi_c$ and $\psi_a$, which denote the equilibrium and transient CDW orders, respectively. As we show in Supplementary Note~\ref{sn:GLtheory}, in the presence of phase competition, the minimum-energy solution near a defect core in $\psi_c$ yields a nonzero $\psi_a$. In addition, we find that the characteristic length scale of the transient CDW, $\lambda_a$, can extend well beyond the confines of the defect core, $\lambda_c$ (Fig.\,\ref{fig:cartoon}(b)). In particular, the ratio of $\lambda_a/\lambda_c$ can become large if the normal-state anisotropy between $a$- and $c$-axis is small, which makes the observation of the transient CDW possible even though the defects may be local. 

This work provides an illustration of the kind of phenomena that can be observed in far-from-equilibrium systems where phase competition plays a significant role in determining material properties. We expect the mechanism of seeing competing states near topological defects to be general, and that other ordered states of matter will exhibit a similar phenomenology under the influence of photoexcitation. Not only does this result provide a path forward to discovering other states of matter in the presence of phase competition, it also paves the way for the manipulation and control of other ordered phases with light.

\begin{acknowledgments}
We thank P.A. Lee, E. Demler, B.V. Fine and A. Aristova for illuminating discussions regarding this work. We thank B. Freelon for pioneering the instrumentation work of the keV UED setup at MIT. We acknowledge support from the U.S. Department of Energy, BES DMSE (keV UED), from the Gordon and Betty Moore Foundation’s EPiQS Initiative grant GBMF4540 (data analysis, manuscript writing) and the Skoltech NGP  Program  (Skoltech-MIT  joint  project)  (theory). We acknowledge support from the U.S. Department of Energy BES SUF Division Accelerator \& Detector R\&D program, the LCLS Facility, and SLAC under contract No.'s DE- AC02-05-CH11231 and DE-AC02-76SF00515 (MeV UED at SLAC). Sample growth and characterization work at Stanford was supported by the U.S. Department of Energy, Office of Basic Energy Sciences, under contract number DEAC02-76SF00515. I.-C.T. and H. W. acknowledge support from the U.S. Department of Energy, Office of Science, Office of Basic Energy Sciences, Materials Sciences and Engineering Division, under contract No. DE-SC0012509. Y.-Q.B., X.W., Y.Y., and P.J.-H. acknowledge support from the Center for Excitonics, an Energy Frontier Research Center funded by the U.S. Department of Energy, Office of Science, Office of Basic Energy Sciences, under award number DESC0001088, as well as the Gordon and Betty Moore Foundation’s EPiQS Initiative through grant GBMF4541 (sample preparation and characterization).
\end{acknowledgments}

\section{Methods}

\noindent\textbf{Sample preparation}. Single crystals of LaTe$_3$ were grown by slow cooling of a binary melt \cite{Ru2006}. Samples were prepared via mechanical exfoliation down to a thickness $\leq60$\,nm, as characterized by atomic force microscopy measurements. Thin flakes were transferred to a commercial 10-nm-thick silicon nitride window (SiMPore Inc.), which was mounted on a copper sample card for UED measurement. All preparations were performed in an inert gas environment as $R$Te$_3$ compounds are known to degrade in air \cite{Ru2006}.

\noindent\textbf{MeV ultrafast electron diffraction}. The experiments were carried out in the Accelerator Structure Test Area facility at SLAC National Laboratory \cite{Weathersby2015,Shen2018}. The 800-nm (1.55-eV), 80-fs pump pulse from a commercial Ti:sapphire regenerative amplifier (RA) laser (Vitara and Legend Elite HE, Coherent Inc.) were focused to an area larger than $500\times500$\,$\upmu$m$^2$ (FWHM) in the sample at an incidence angle around $5^{\circ}$ from sample normal. 3.1\,MeV electron bunches were generated by radio-frequency photoinjectors at a repetition rate of 180\,Hz. The electron beam was normally incident on the sample with a $90\times90$\,$\upmu$m$^2$ (FWHM) spot size. The laser and electron pulses were spatially overlapped on the sample, and their relative arrival time was adjusted by a linear translation stage. The diffraction pattern was imaged by a phosphor screen (P-43) and recorded by an electron-multiplying charge-coupled device (EMCCD, Andor iXon Ultra 888). A circular through hole in the center of the phosphor screen allowed the passage of undiffracted electron beam to prevent camera saturation. Samples were maintained at 307\,K during the measurement. The overall temporal resolution is around 300\,fs, as determined via electron streaking by an intense, single-cycle terahertz pulse \cite{DSD2019}.

\noindent\textbf{keV ultrafast electron diffraction}. The light-induced CDW was reproduced in a separate UED setup at MIT with a different pump pulse wavelength and probe electron kinetic energy (see Supplementary Note~\ref{sn:keV}). The setup adopts a compact geometry \cite{Zong2019}. The 1038\,nm (1.19\,eV), 190\,fs output of a Yb:KGW RA laser system (PHAROS SP-10-600-PP, Light Conversion) was focused to a $500\times500$\,$\upmu$m$^2$ (FWHM) area in the sample. The electron beam was generated by focusing the fourth harmonic (260\,nm, 4.78\,eV) to a gold-coated sapphire photocathode in high vacuum ($<4\times10^{-9}$\,torr). Photoelectrons excited were accelerated to 26\,kV in a dc field and focused to an aluminum-coated phosphor screen (P-46) by a magnetic lens, with a $270\times270$\,$\upmu$m$^2$ (FWHM) beam spot at the sample position. Diffraction patterns were recorded by a commercial intensified CCD (iCCD PI-MAX II, Princeton Instruments). The laser repetition rate used was 10\,kHz, and the operating temporal resolution is about 1\,ps, as determined from the initial response of the CDW peak intensity \cite{Zong2019}. Measurements in this setup were performed at room temperature.

\bibliography{lib}

\newpage
\beginsupplement
\onecolumngrid
\begin{center}
\textbf{\large Supplementary Information for ``Light-Induced Charge Density Wave in LaTe$_\mathbf{3}$''}
\end{center}\hfill\break
\twocolumngrid

\rSN{sn:keV}
\subsection{Supplementary Note~\ref{sn:keV}: Detection of light-induced CDW by both MeV and keV UED setups}

\begin{figure*}[htb!]
	\includegraphics[width=1.0\textwidth]{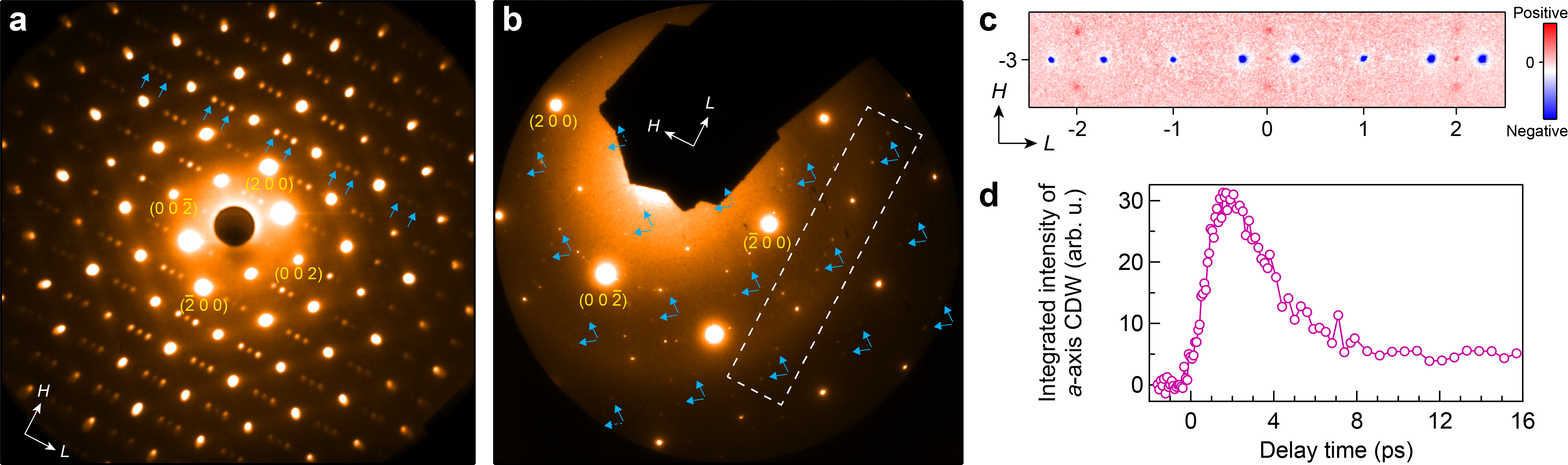}
	\caption{\textbf{a-b},~Full diffraction patterns of LaTe$_3$ before photoexcitation, taken with 3.1\,MeV and 26\,keV electron kinetic energy, respectively. ($\pm2$~0~0) and (0~0~$\pm2$) peaks are labeled. Blue arrows indicate examples of equilibrium $c$-axis CDW peaks. Dashed rectangle in \textbf{b} denotes the region of interest examined in \textbf{c}. In both images, the undiffracted central electron beam is omitted to prevent camera saturation. \textbf{c},~Differential intensity plot from keV UED measurements, focusing on the $H=-3$ row in \textbf{b} and showing photo-induced change at 1.5\,ps with respect to $-1.3$\,ps. \textbf{d},~Time evolution of integrated intensities of the transient $a$-axis CDW peak from keV UED. Peaks in multiple Brillouin zones are averaged for improved signal-to-noise ratio. The curve is vertically offset so values before photoexcitation are averaged to zero. The incident pump laser fluence for \textbf{c-d} was 240\,$\upmu$J/cm$^{2}$.}
\label{fig:keV}
\end{figure*}

The photo-induced $a$-axis CDW peaks are observed in both MeV and keV UED setups. Compared to MeV electron diffraction, keV diffraction possesses significantly improved momentum resolution, but suffers more background scattering from the 10-nm-thick silicon nitride substrate (compare Fig.\,\ref{fig:keV}(a) and (b)), which makes the detection of weak intensities from the transient $a$-axis CDW more challenging. Nonetheless, the differential diffraction plot reveals clear peaks along the $a$-axis at 1.5\,ps after photoexcitation, while the intensity of the equilibrium $c$-axis CDW peaks is suppressed (Fig.\,\ref{fig:keV}(c)). Tracking the intensity of the $a$-axis CDW peaks over various pump-probe delays in the keV UED dataset (Fig.\,\ref{fig:keV}(d)), we reproduce a similar temporal evolution characterized by a transient enhancement followed by disappearance over a few picoseconds. It is worth noting that the pump laser wavelengths are different in the two setups (800\,nm and 1038\,nm, see Methods), so the observed light-induced CDW is unlikely a result of any resonantly-pumped inter-band transition.

\rSN{sn:rise_time}
\subsection{Supplementary Note~\ref{sn:rise_time}: Initial system response after photoexcitation}

In the main text, we noted that the initial decay of the $c$-axis CDW occurs faster than the rise of the transient $a$-axis CDW. In this section, we offer two remarks on this difference.

\begin{figure*}[htb!]
	\includegraphics[scale=0.56]{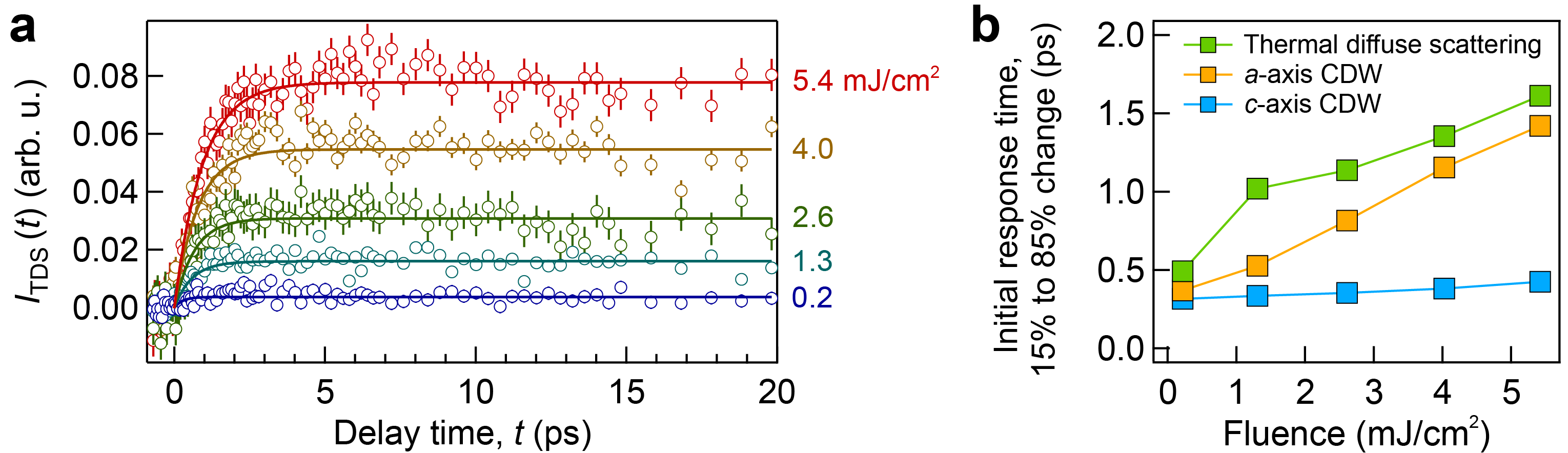}
	\caption{\textbf{a},~Time evolution of thermal diffuse scattering intensities, integrated over the green circles in Fig.\,\ref{fig:timetrace}(c). Each color denotes an incident fluence. Error bars are obtained from the standard deviation of noise prior to photoexcitation. Curves are single-exponential fits for $t\geq0$. \textbf{b},~Initial response time for the rise of thermal diffuse scattering (green), the rise of the transient $a$-axis CDW peak (orange), and the suppression of the equilibrium $c$-axis CDW peak (blue). For consistency, the characteristic time is taken as the interval between 15\% to 85\% of the initial change shown in \textbf{a} and in Fig.\,\ref{fig:fludepend}(a-b).}
\label{fig:tds}
\end{figure*}

First, the microscopic processes underlying the light-induced CDW suppression along the $c$-direction and the formation along the $a$-direction are different, so the two timescales do not have to match. Specifically, the suppression involves the displacive excitation of the coherent amplitude mode phonon \cite{Hellmann2012, Schmitt2008,Rettig2016}, so the timescale is set by the amplitude mode period if one neglects more subtle effects such as dynamical slowing down \cite{Yusupov2010,DSD2019}. On the other hand, there is no coherent motion of lattice ions when the $a$-axis CDW order starts to emerge, and longer time is needed for local fluctuations to acquire sufficient phase coherence. Combining this picture with our proposed mechanism that topological defects induced in the $c$-axis CDW give rise to the transient $a$-axis CDW, we note that the time to form the $a$-axis CDW is necessarily longer than the time to suppress the $c$-axis CDW. This is indeed the case for all pump laser fluences investigated, as shown in Fig.\,\ref{fig:tds}(b).

Second, the apparent slow rise in the $a$-axis CDW peak intensity can be partially attributed to the slow rise in the thermal diffuse scattering intensity. In Fig.\,\ref{fig:tds}(a), we plot the time evolution of thermal diffuse scattering at various fluences, and summarize its characteristic rise time in Fig.\,\ref{fig:tds}(b) (green squares). As noted earlier in the main text, the intensity of the $a$-axis CDW peak, $I_a$, has a significant contribution from thermal diffuse scattering, which, for example, gives a non-zero plateau to $I_a$ at long time delay (Figs.\,\ref{fig:timetrace}(b) and \ref{fig:fludepend}(a)). Therefore, the initial response time in the $I_a$ trace is necessarily affected by the time evolution of thermal diffuse scattering as well.

\rSN{sn:q_cdw}
\subsection{Supplementary Note~\ref{sn:q_cdw}: Wavevectors of equilibrium and transient CDWs}

In Fig.\,\ref{fig:cartoon}(a), we summarized the wavevectors of equilibrium $c$- and $a$-axis CDW, $q_c$ and $q_a$, and compared them to the wavevector of the photo-induced CDW, $\widetilde{q}_a$. In this section, we discuss how we determine the wavevector from UED experiments and comment on the trends of $q_{a,c}$ across different temperatures or across rare-earth elements.

\subsubsection{\ref{sn:q_cdw}.1 Measuring CDW wavevectors from electron diffraction}

Despite the poor momentum resolution of MeV electron diffraction due to the short de~Broglie wavelength ($\lambda = 0.35$\,pm for electrons with 3.1\,MeV kinetic energy) \cite{Weathersby2015}, the large number of peaks observed across multiple Brillouin zones can yield a precise value of the CDW wavevector after statistical averaging. As the CDW wavevector is expressed in terms of the reciprocal lattice unit (r.l.u.), we measure the distance between an adjacent pair of CDW peaks in the diffraction pattern, normalized by the distance between the neighboring Bragg peaks. This normalization procedure minimizes distortion of the wavevector due to any misalignment of the electron beam to the sample surface normal, or due to the slight curvature of the Ewald sphere.

\begin{figure*}[htb!]
	\includegraphics[scale=0.56]{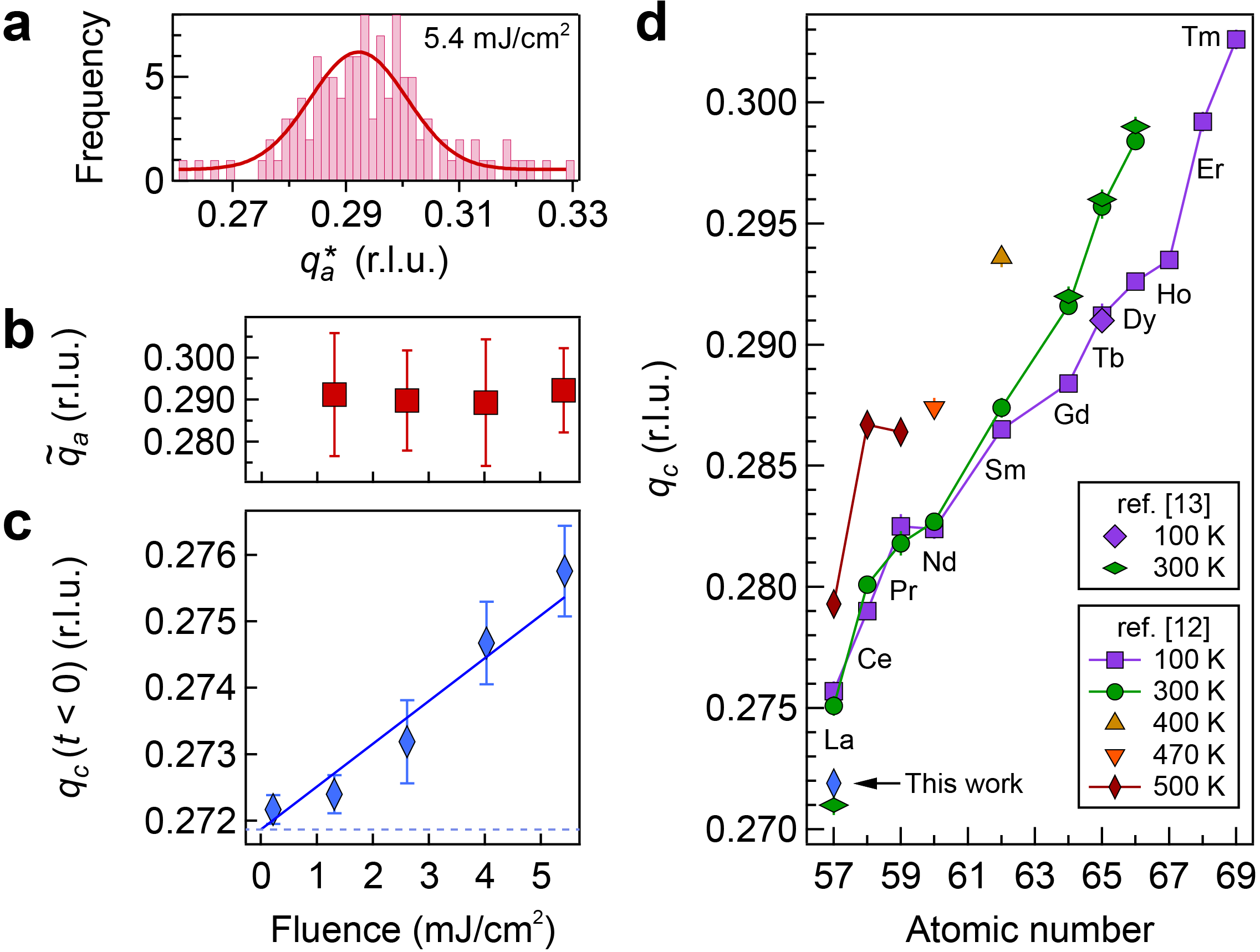}
	\caption{\textbf{a},~Histogram of wavevector $\widetilde{q}_a$ of 114 pairs of transient $a$-axis CDW peaks taken between time delays from 0.8\,ps to 10.4\,ps at 5.4\,mJ/cm$^2$ pump laser fluence. Curve is a Gaussian fit, whose center and width determine the value and uncertainty of $\widetilde{q}_a$. \textbf{b},~Wavevector $\widetilde{q}_a$ of the transient CDW at different pump fluences. Error bars represent the half-width at half maximum of the Gaussian fit at each fluence; see \textbf{a}. The $a$-axis CDW peak is too weak at 0.2\,mJ/cm$^2$ to have its wavevector determined reliably. \textbf{c},~Wavevector $q_c$ of the equilibrium $c$-axis CDW peak taken before the arrival of the pump laser pulse at various fluences. Steady-state laser heating causes $q_c$ to increase monotonically with fluence, consistent with previous reports \cite{Ru2008,Malliakas2006,Banerjee2013}. Error bars represent one standard deviation among multiple peaks selected. The equilibrium value of $q_c$ at the base temperature of 307\,K is determined from the vertical intercept (dashed line) of the linear fit (solid line) at zero fluence. \textbf{d},~Wavevector $q_c$ of the equilibrium CDW at various temperatures. The value of $q_c$ determined from \textbf{c} (blue diamond) is benchmarked against previous reports by Ru \textit{et al.} \cite{Ru2008thesis} and Malliakas \textit{et al.} \cite{Malliakas2006}.}
\label{fig:q_cdw}
\end{figure*}

To determine $\widetilde{q}_a$ of the transient $a$-axis CDW, we apply the above procedure at each time delay when the photo-induced peak can be distinguished from the background. The location of each peak is determined by fitting it to a Gaussian profile, and $\widetilde{q}_a$ is computed for each adjacent pair. We find no observable time-dependence of $\widetilde{q}_a$ beyond our experimental uncertainty. Hence, for each fluence, we plot a histogram of all $\widetilde{q}_a$ extracted, fit it to a Gaussian distribution, and assign the Gaussian center and width as the value and uncertainty of $\widetilde{q}_a$; see Fig.\,\ref{fig:q_cdw}(a) for an example at 5.4\,mJ/cm$^2$ incident fluence. Figure~\ref{fig:q_cdw}(b) shows the fluence dependence of $\widetilde{q}_a$, which displays a constant trend within the uncertainties. The value (or uncertainty) of $\widetilde{q}_a$ quoted in Fig.\,\ref{fig:cartoon}(a) (red square) is thus taken as the average of the values (or uncertainties) shown in Fig.\,\ref{fig:q_cdw}(b). It is worth noting the consistent value of $\widetilde{q}_a$ measured from the keV UED experiment (Fig.\,\ref{fig:cartoon}(a), orange square), which has a reduced error bar due to improved momentum resolution benefiting from a much longer de~Broglie wavelength ($\lambda=7.5$\,pm for electrons with 26\,keV kinetic energy). In this case, $\widetilde{q}_a$ is computed from the statistical average of CDW pairs from 14 different Brillouin zones in the differential diffraction plot; the $H=-3$ row is shown in Fig.\,\ref{fig:keV}(c).

To confirm the validity of the above procedure of statistical averaging and computing $\widetilde{q}_a$, we apply a similar method to measure the wavevector of the equilibrium $c$-axis CDW, $q_c$. As $q_c$ is known to change after photoexcitation \cite{Moore2016}, we focus on the diffraction images at delay time $t<0$. Figure~\ref{fig:q_cdw}(c) summarizes $q_c(t<0)$ for each laser fluence, where steady-state laser heating causes $q_c$ to increase monotonically with fluence, consistent with previous reports \cite{Ru2008,Malliakas2006,Banerjee2013}. To determine $q_c$ at equilibrium, we extrapolate its value at zero incident fluence through a linear fit (Fig.\,\ref{fig:q_cdw}(c)), which is consistent with values obtained in high-resolution X-ray measurement at a similar temperature (Fig.\,\ref{fig:q_cdw}(d)). In particular, the value obtained by our method (blue diamond) is in nearly perfect agreement with the value reported by N.~Ru (green diamond) \cite{Ru2008thesis}, but is slightly smaller than that from C.~D.~Malliakas and co-workers (green circle) \cite{Malliakas2006}. This difference may arise from different crystal growth methods; the LaTe$_3$ crystals used in the present work is grown by the same procedure as in N.~Ru's study \cite{Ru2006,Ru2008thesis}.

\subsubsection{\ref{sn:q_cdw}.2 Temperature dependence of CDW wavevectors}

As alluded to earlier and illustrated in Fig.\,\ref{fig:q_cdw}(c), the CDW wavevector at equilibrium varies with temperature. The variation is significant for $q_c$, and an example is given in Fig.\,\ref{fig:q_cdw}(d) for different $R$Te$_3$. Hence, to draw the trend of $q_c$ across the $R$Te$_3$ series in Fig.\,\ref{fig:cartoon}(a), for fair comparison, we use the values of $q_c$ reported by C.~D.~Malliakas \textit{et al.} \cite{Malliakas2006} that are closest to $T_{c1}$. We note that the exact temperature is not essential to the observed trend because a similar one can be seen at other temperatures in Fig.\,\ref{fig:q_cdw}(d) as well.

\begin{figure*}[htb!]
	\includegraphics[scale=0.56]{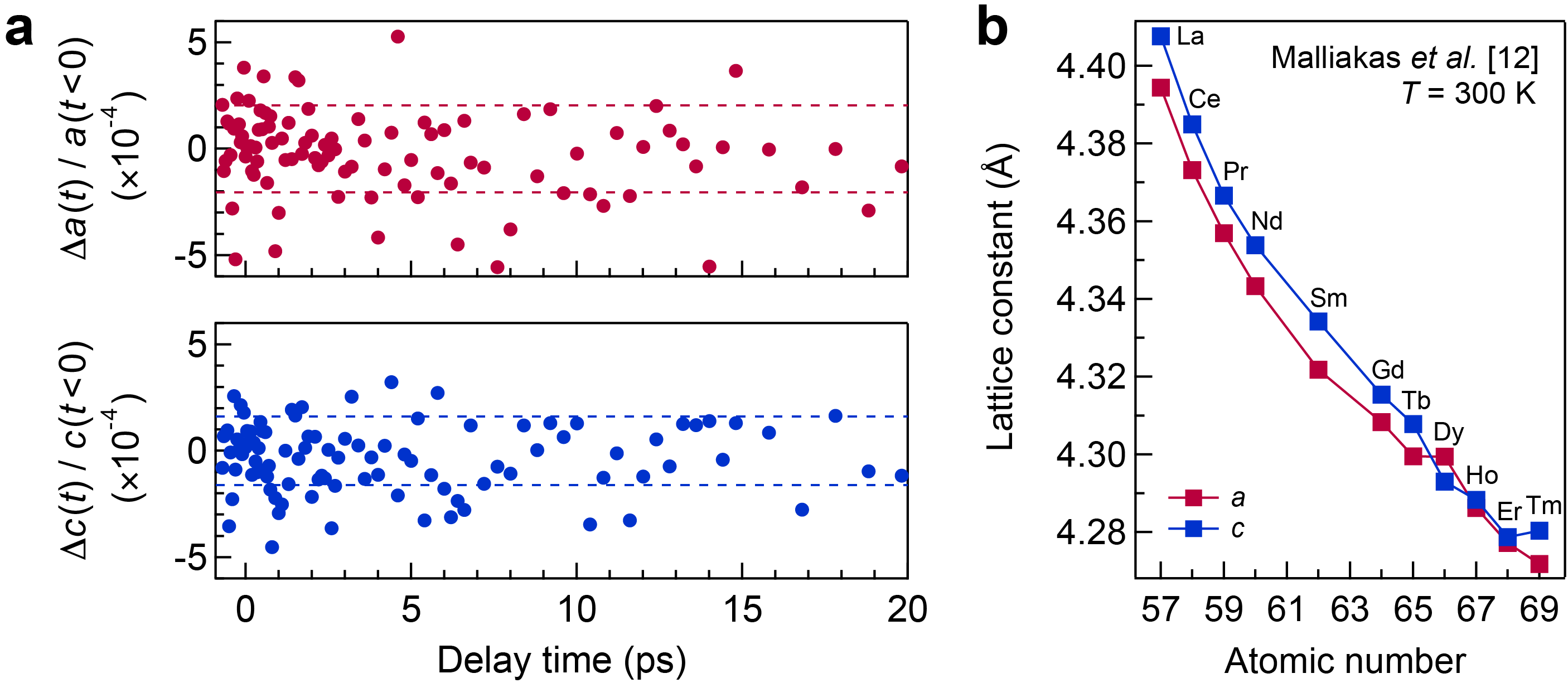}
	\caption{\textbf{a},~Time evolution of normalized \textit{change} in lattice constants along the $a$-axis (upper panel) and $c$-axis (lower panel) at 5.4\,mJ/cm$^2$ incident fluence. Dashed lines demarcate one standard deviation for data points across all time delays, indicating no observable time-dependent change in either lattice constant beyond 0.02\%. \textbf{b},~Lattice constants, $a$ and $c$, for different $R$Te$_3$, taken from X-ray measurements at 300\,K by Malliakas \textit{et al.} \cite{Malliakas2006}. Error bars are smaller than the symbol size.}
\label{fig:ac_lattice}
\end{figure*}

In the main text and in Fig.\,\ref{fig:cartoon}(a), we noted that $q_{a,\text{soft}}\approx q_c$ and $\widetilde{q}_a \approx q_c$. While the former relationship looks exact in the example of DyTe$_3$ (white square in Fig.\,\ref{fig:cartoon}(a)), the latter shows slightly worse agreement in LaTe$_3$ (orange and red squares in Fig.\,\ref{fig:cartoon}(a)). This is because the $q_c$ value quoted for DyTe$_3$ is at 300\,K, close to its $T_{c1}=306(3)$\,K \cite{Ru2008}. On the other hand, $q_c$ for LaTe$_3$ is quoted at 500\,K, still considerably lower than its projected $T_{c1}=670$\,K \cite{Hu2014}. From the temperature trend in Fig.\,\ref{fig:q_cdw}(d), we speculate that the relationship $\widetilde{q}_a \approx q_c(T=T_{c1})$ is also exact.

Compared to $q_c$, the value of $q_a$ at equilibrium has a much smaller temperature dependence. For example, $q_a$ varies by less than 0.07\% from $0.84T_{c2}$ to $0.18T_{c2}$ in TbTe$_3$, about 10 times smaller than the variation in $q_c$ across the same temperature range in terms of $T_{c1}$ \cite{Banerjee2013}.

\subsubsection{\ref{sn:q_cdw}.3 Remarks on wavevector trends across the \textit{R}Te$_\mathit{3}$ series}

If one compares $q_a$ and $q_c$ across the $R$Te$_3$ series, as highlighted in Fig.\,\ref{fig:cartoon}(a), they possess opposite trends. In particular, their values become more similar towards heavier rare-earth elements, which is reminiscent of the trends in $T_{c1}$ and $T_{c2}$ shown in Fig.\,\ref{fig:intro}(b). The opposite trends in $q_a$ and $q_c$ are another manifestation of the two competing CDW orders tuned by the chemical pressure from the rare-earth ions. Below we provide two different ways to understand these trends.

First, from systematic band structure mapping by angle-resolved photoemission spectroscopy (ARPES) and simple analysis from a tight-binding model \cite{Brouet2008}, it was shown that $q_c$ anti-correlates with $\Delta_c$ and hence with $T_{c1}$, where $\Delta_c$ is the maximum gap size due to the formation of the $c$-axis CDW. By the same argument, one would expect a similar anti-correlation between $q_a$ and $T_{c2}$ if subtle details of Fermi surface reconstruction below $T_{c1}$ are neglected.

Second, the observation that $q_{a,\text{soft}}\approx q_c$ but $q_a>q_c$ at equilibrium (Fig.\,\ref{fig:cartoon}(a)) is indicative of the Fermi surface anisotropy introduced by the opening of the $c$-axis CDW gap. Specifically, when the low-temperature CDW forms along the $a$-axis, the anisotropy results in $q_a(T=T_{c2}) > q_{a,\text{soft}}(T=T_{c1})$. As we move from heavier to lighter rare-earth element (Tm to La), one would expect the anisotropy to grow due to an increasing $c$-axis CDW gap size. Correspondingly, one would expect $q_a$ to increase from Tm to La as well.

\rSN{sn:other_theories}
\subsection{Supplementary Note~\ref{sn:other_theories}: Alternative mechanisms for the light-induced CDW}

Here, we discuss two viable alternative scenarios that could lead to the existence of the out-of-equilibrium charge density wave. It is also worth noting that the interpretation may be complicated by large phase fluctuations, which are difficult to distinguish from topological defects.

\subsubsection{\ref{sn:other_theories}.1 Light-induced strain}

It is well understood that the CDW properties in the $R$Te$_3$ family is significantly modified by the chemical pressure exerted by the rare-earth ion, leading to systematic trends in $T_{c1}$ and $T_{c2}$ (Fig.\,\ref{fig:intro}(b)). Across the series, the lattice constants, $a$ and $c$, also change systematically (more than 2\% from La to Tm) due to the changing size of the rare-earth ions (Fig.\,\ref{fig:ac_lattice}(b)). Therefore, it is conceivable that photoexcitation could transiently strain the crystal, which would cause the lattice constants to change and induce the $a$-axis CDW. Furthermore, it may be the case that such strain can lead to a reduction in the $a$/$c$ lattice anisotropy, which could also lead to a preference of the $a$-axis CDW over the $c$-axis CDW.

To test this scenario, we examined the change in the lattice parameters as a function of time after photoexcitation by tracking the position of lattice Bragg peaks. In Fig.\,\ref{fig:ac_lattice}(a), we plot the in-plane lattice parameters as a function of pump-probe delay time. These plots demonstrate no observable change in the lattice parameters beyond the experimental uncertainty level of 0.02\%. The scenario of the non-equilibrium CDW induced by strain must thus be ruled out.

\subsubsection{\ref{sn:other_theories}.2 Kibble-Zurek-like domains}

\noindent\textbf{Global melting}. The Kibble-Zurek mechanism (KZM) describes a phase transition involving a thermal quench, which gives rise to topological defects, possibly in the form of domain walls \cite{Kibble1976,Zurek1996}. Considering this scheme, it is plausible that upon photoexcitation, the equilibrium CDW completely vanishes, and during relaxation, some regions of the sample relax into an $a$-axis CDW while others relax into a $c$-axis CDW in a probabilistic fashion, resulting in the formation of domains. Following this initial relaxation process, the $c$-axis domains would then engulf the $a$-axis domains over the observed relaxation timescale. 

While this scenario is worth considering, there is a clear observation that is antithetical to this picture. Even for small fluences, the $a$-axis CDW grows in intensity, as evidenced in Fig.\,\ref{fig:fludepend}(a). This observation suggests that it is not necessary to fully melt the $c$-axis CDW to create the $a$-axis CDW.

\noindent\textbf{Local melting}. One may argue that the photons could locally melt the CDW in ``patches", and within those patches, there could be a probabilistic growth of $a$- or $c$-axis CDW domains. Although we cannot rule out this scenario definitively, the time-resolved ARPES data in LaTe$_3$ from Ref.~\cite{Zong2019} does not show relaxation timescales consistent with this picture. If local melting was occurring, one would expect two relaxation timescales in the ARPES data: first for a development of the CDW amplitude in the locally melted regions, and second for the subsequent engulfing of the $a$-axis domains by the $c$-axis CDW. However, these two timescales were not reliably observed in the time-resolved ARPES measurements of Ref.~\cite{Zong2019}. Thus, while the local melting scenario remains a possibility in explaining the observed non-equilibrium CDW, we find the topological defect picture more likely.

\rSN{sn:GLtheory}
\subsection{Supplementary Note~\ref{sn:GLtheory}: Ginzburg-Landau formalism of two competing orders}

\subsubsection{\ref{sn:GLtheory}.1 Ginzburg-Landau free energy}

To describe the properties of the competition between the two CDW order parameters, $\psi_a$ and $\psi_c$, we introduce the following Ginzburg-Landau free energy density in two spatial dimensions:
\begin{eqnarray}
\label{eq:free_energy_LG}
    \mathcal{W}&=& \left(r_c|\psi_c|^2 + \frac{\beta_c|\psi_c|^4}{2}
    + \kappa_c|\nabla_\mathbf{r}\psi_c|^2\right) \\
    &+&\left( r_a|\psi_a|^2 + \frac{\beta_a|\psi_a|^4}{2} + \kappa_a|\nabla_\mathbf{r}\psi_a|^2\right)
    + \eta|\psi_c|^2|\psi_a|^2.\notag
\end{eqnarray}
Here $r_{i=c,a},\, \beta_i,\, \kappa_i$, and $\eta$ are the model parameters. The last term describes the competition between the two phases. We choose this term to be quadratic in both orders for two reasons. First, such a term is allowed by symmetry, and second, it is the lowest order term that would give rise to the observed competition. Therefore, this is the simplest model capturing all the essential physics. The mean-field phase diagram of the above model, Eq.\,\eqref{eq:free_energy_LG},
has two regimes \cite{chaikin1995principles}: for $\beta_a \beta_c > \eta^2$ (tetracritical regime), the two phases can coexist ($\psi_a \neq 0$, $\psi_c \neq 0$); for $\beta_a \beta_c < \eta^2$ (bicritical regime), only one phase may develop ($\psi_c \neq 0$, $\psi_a=0$). Given the evolution of the CDWs with rare earth mass in $R$Te$_3$, we assume that the bicritical regime applies for LaTe$_3$, as only one CDW exists in equilibrium. On the other hand, the tetracritical case may describe the tritellurides where both $a$- and $c$-axis CDWs exist at finite temperature (Fig.\,\ref{fig:intro}(b)).

An alternative point of view for the bicritical regime in LaTe$_3$ is as follows: due to the small $a/c$-anisotropy, we expect that the unrenormalized transition temperature of the sub-dominant order, $T^*_{c2}$, is close to the observed transition temperature of the dominant order, $T_{c1}$, with $T_{c1} \gtrsim T^*_{c2}$. Here, $r_a = A_a(T-T^*_{c2})$ and $r_c = A_c(T-T^*_{c1})$, where $A_a$ and $A_c$ are positive constants. At the simple mean-field level, $T^*_{c1} = T_{c1}$: since the dominant order sets in first, its transition temperature is not renormalized by the sub-dominant order. On the other hand, the renormalized transition temperature for the sub-dominant order, $T_{c2}$, which would be observed in experiments, can be suppressed to a negative value, since $T_{c2} = T^*_{c2} - \eta |\psi_c(T_{c2})|^2/A_a$. This is likely the case for LaTe$_3$, where the $a$-axis CDW does not appear in equilibrium. 

We anticipate that thermal fluctuations of both order parameters, which are neglected in the mean-field treatment, will have a strong impact on the equilibrium phase diagram. In this regard, the effect of the sub-dominant CDW fluctuations on the ground state properties are expected to be profound, especially between temperatures $T^*_{c2}$ and $T_{c2}$. We also expect a strong influence on the actual transition temperature of the dominant CDW, $T_{c1}$.

\subsubsection{\ref{sn:GLtheory}.2 Vortex solution}

In the main text, we presented a consistent interpretation of the experimental results. Photoexcitation creates topological defects in the dominant order parameter $\psi_c$, and at the cores of these defects, the competing phase $\psi_a$ can develop. Below we study the properties of these two competing orders near a topological defect in $\psi_c$.

Minimizing the total free energy, $\displaystyle\int d^2\mathbf{r}\, \mathcal{W}(\mathbf{r})$, one obtains the following equations:
\begin{eqnarray}
-\kappa_c \nabla_\mathbf{r}^2 \psi_c + r_c \psi_c + \beta_c |\psi_c|^2 \psi_c + \eta |\psi_a|^2\psi_c  &= 0,\label{eqn:psi_c}\\
-\kappa_a \nabla_\mathbf{r}^2 \psi_a + r_a \psi_a + \beta_a |\psi_a|^2 \psi_a + \eta |\psi_c|^2\psi_a  &= 0.\label{eqn:psi_a}
\end{eqnarray}
Below we assume that $\kappa_a = \kappa_c = \kappa$, $\beta_a = \beta_b = \beta$, $r_c < r_a < 0$ and $\eta^2 \geq \beta^2$. The latter two conditions imply the bicritical regime, where only $\psi_c$ develops. The first two conditions are justified by the small $a/c$-anisotropy; the two orders are only differentiated by $r_a$ and $r_c$. We emphasize that our subsequent analysis can be easily generalized to the tetracritical regime, where the main conclusions do not change.

Assuming the photo-induced topological defect in the $c$-axis CDW takes the form of a vortex (i.e. a CDW dislocation), similar to the observation in Ref.~\cite{Fang2019}, we seek a vortex solution in $\psi_c$ and solve for $\psi_a$ in a self-consistent way. In the cylindrical coordinates $(r, \phi)$ where the vortex is located at $r=0$, the solution has the form: $\psi_c(r, \phi) = \psi_c^{\infty} f(r) e^{im\phi}$, $\psi_a(r, \phi) = \psi_a^{\infty} g(r)$, where $\psi_{i=c,a}^{\infty} = \sqrt{-r_i/\beta}$ and $m = \pm 1,\pm 2,\dots$ is the vorticity of the vortex. $f(r)$ and $g(r)$ are smooth functions of $r$, representing normalized order parameters. Then, Eqs.\,\eqref{eqn:psi_c} and \eqref{eqn:psi_a} take the following form:
\begin{eqnarray}
&&\xi_c^2 (f'' + \frac{1}{r}f') = \frac{\xi^2_c m^2}{r^2}f - f + f^3 + \alpha_c f g^2,\label{eqn:f_vortex}\\
&&\xi_a^2(g'' + \frac{1}{r}g') = -g + g^3 + \alpha_a f^2 g,\label{eqn:g_vortex}
\end{eqnarray}
where $\displaystyle\xi_{i=c,a}^2 =- \frac{\kappa}{r_i},\, \alpha_a = \frac{\eta}{\beta}\frac{r_c}{r_a} > 1$, and $\displaystyle\alpha_c = \frac{\eta}{\beta}\frac{r_a}{r_c} > 0$. From these equations, we can calculate the asymptotic behaviour of the order parameters: 
\begin{eqnarray}
&&\begin{cases}
f_{r\rightarrow \infty} = 1 - C_1  \exp{(-\sqrt{2}\,r/\xi_c)},\\
f_{r\rightarrow 0} = C_3 r^{|m|},
\end{cases}\\
&&\begin{cases}
g_{r\rightarrow \infty} =  C_2 \exp{(-\sqrt{\alpha_a-1}\,r/\xi_a)},\\ 
g_{r\rightarrow 0} = C_4 + C_5 r^2.
\end{cases}
\end{eqnarray}
Here $C_i$ are constants that may be obtained by numerically solving Eqs.\,\eqref{eqn:f_vortex} and \eqref{eqn:g_vortex}. From these analyses we learn that
\begin{equation}
\frac{\lambda_a}{\lambda_c} = \frac{\xi_a}{\xi_c} \sqrt{\frac{2}{\alpha_a - 1}} = \sqrt{\frac{\frac{2 r_c}{r_a}}{\frac{\eta}{\beta}\frac{r_c}{r_a} - 1}},
\end{equation}
where $\lambda_a = \xi_a/\sqrt{\alpha_a - 1}$ is the spatial extent of the sub-dominant phase $\psi_a$, and $\lambda_c = \xi_c/\sqrt{2}$ is the characteristic length scale of the defect core in $\psi_c$ (Fig.\,\ref{fig:cartoon}(b)). Notably, when the $a$-axis and $c$-axis become more isotropic ($r_c/r_a\rightarrow1+$), the ratio $\lambda_a/\lambda_c$ becomes larger. In the case of $R$Te$_3$ where $a/c$-anisotropy is small, the relatively large spatial extent of $\psi_a$ would make its observation easier in a diffraction experiment.

\end{document}